\documentclass[twocolumn]{aastex7}

\def\feka{Fe~K$\alpha$}
\def\fekb{Fe~K$\beta$}

\begin{document}

\title{The dynamic central environment of NGC~3516 revealed by XRISM}

\author[orcid=0000-0002-7292-6852,gname='Anna',sname='Jur\'{a}\v{n}ov\'{a}']{Anna Jur\'{a}\v{n}ov\'{a}}
\affiliation{MIT Kavli Institute for Astrophysics and Space Research, Massachusetts Institute of Technology, Cambridge, MA 02139, USA}
\email[show]{ajuran@mit.edu} 

\author[orcid=0000-0003-0172-0854]{Erin~Kara} 
\affiliation{Kavli Institute for Astrophysics and Space Research, Massachusetts Institute of Technology, MA 02139, USA}
\email{ekara@mit.edu}

\author[orcid=0000-0001-9735-4873]{Ehud~Behar}
\affiliation{Department of Physics, Technion, Technion City, Haifa 3200003, Israel}
\email{behar@physics.technion.ac.il}

\author[0000-0001-8470-749X]{Elisa~Costantini}
\affiliation{SRON Netherlands Institute for Space Research, Leiden, The Netherlands}
\affiliation{Anton Pannekoek Institute for Astronomy, University of Amsterdam, Science Park 904, NL-1098 XH Amsterdam, The Netherlands}
\email{e.costantini@sron.nl}

\author[orcid=0000-0003-2869-7682]{Jon~M.~Miller}
\affiliation{Department of Astronomy, University of Michigan, MI 48109, USA}
\email{jonmm@umich.edu}

\author[orcid=0000-0002-0786-7307]{Daniele~Rogantini}
\affiliation{Department of Astronomy and Astrophysics, University of Chicago, 5640 S Ellis Ave, Chicago, IL 60637, USA}
\email{danieler@uchicago.edu}

\author[orcid=0000-0002-0982-0561]{James~N. Reeves}
\affiliation{Department of Physics, Institute for Astrophysics and Computational Sciences, The Catholic University of America, Washington, DC 20064, USA}
\affiliation{INAF, Osservatorio Astronomico di Brera, Via Bianchi 46, I-23807 Merate (LC), Italy}
\email{reevesjn@cua.edu}

\author[orcid=0000-0002-2629-4989]{Valentina~Braito}
\affiliation{INAF, Osservatorio Astronomico di Brera, Via Bianchi 46, I-23807 Merate (LC), Italy}
\affiliation{Department of Physics, Institute for Astrophysics and Computational Sciences, The Catholic University of America, 620 Michigan Ave., N.E., Washington, DC 20064, USA}
\affiliation{Dipartimento di Fisica, Universit\`a di Trento, Via Sommarive 14, I-38123, Trento, Italy}
\email{valentina.braito@inaf.it}

\author[orcid=0000-0001-5924-8818]{Jacobo~Ebrero}
\affiliation{Telespazio UK for the European Space Agency (ESA), European Space Astronomy Centre (ESAC), Camino Bajo del Castillo, s/n, 28692 Villanueva de la Ca\~{n}ada, Madrid, Spain}
\email{Jacobo.ebrero.carrero@ext.esa.int}

\author[0009-0006-4968-7108]{Luigi~Gallo}
\affiliation{Department of Astronomy and Physics, Saint Mary's University, Nova Scotia B3H 3C3, Canada}
\email{lgallo@ap.smu.ca}

\author[0000-0003-2535-6436]{Noa~Keshet}
\affiliation{Department of Physics, Technion, Technion City, Haifa 3200003, Israel}
\email{noa.keshet@campus.technion.ac.il}

\author[orcid=0000-0002-2180-8266]{Gerard~A.~Kriss}
\affiliation{Space Telescope Science Institute, 3700 San Martin Drive, Baltimore, MD 21218, USA}
\email{gak@stsci.edu}

\author[orcid=0000-0002-4992-4664]{Missagh~Mehdipour}
\affiliation{Department of Astronomy, University of Michigan, 1085 South University Avenue, Ann Arbor, MI 48109, USA}
\affiliation{Space Telescope Science Institute, 3700 San Martin Drive, Baltimore, MD 21218, USA}
\email{missagh@umich.edu}

\author[0000-0001-6020-517X]{Hirofumi~Noda}
\affiliation{Astronomical Institute, Tohoku University, Miyagi 980-8578, Japan}
\email{hirofumi.noda@astr.tohoku.ac.jp}

\author[0000-0002-0114-5581]{Atsushi Tanimoto}
\affiliation{Department of Science, Kagoshima University, Kagoshima 890-0065, Japan}
\email{atsushi.tanimoto@sci.kagoshima-u.ac.jp}

\author[orcid=0000-0002-6562-8654]{Francesco~Tombesi}
\affiliation{Physics Department, Tor Vergata University of Rome, Via della Ricerca Scientifica 1, 00133 Rome, Italy}
\affiliation{INAF – Astronomical Observatory of Rome, Via Frascati 33, 00040 Monte Porzio Catone, Italy}
\affiliation{INFN - Rome Tor Vergata, Via della Ricerca Scientifica 1, 00133 Rome, Italy }
\email{francesco.tombesi@roma2.infn.it}

\author[orcid=0000-0003-2971-1722]{Tracey~J.~Turner}
\affiliation{Eureka Scientific, Inc., 2452 Delmer Street Suite 100, Oakland, CA 94602-3017, USA}
\email{turnertjane@gmail.com}

\author[0000-0002-9754-3081]{Satoshi Yamada}
\affiliation{Frontier Research Institute for Interdisciplinary Sciences, Tohoku University, Sendai 980-8578, Japan}
\affiliation{Astronomical Institute, Tohoku University, 6-3 Aramakiazaaoba, Aoba-ku, Sendai, Miyagi 980-8578, Japan}
\email{satoshi.yamada@astr.tohoku.ac.jp}

\begin{abstract}

\noindent We present a detailed, time-resolved analysis of the Fe~K band of the Seyfert~1.5 galaxy NGC~3516 observed with \textit{XRISM}. The 249~ks observation spanning $\sim$310~ks in elapsed time reveals an exceptionally rich and time-variable absorption spectrum. 
Six distinct absorption components are detected across multiple ionization states, spanning more than an order of magnitude in ionization parameter and a wide range of systemic velocities, from a potential inflow ($+4300~\rm km~s^{-1}$) to a mildly relativistic ultra-fast outflow ($-9800~\rm km~s^{-1}$). Despite their diversity, the components exhibit relatively small broadening ($\lesssim$400~km~s$^{-1}$), implying comparable internal dynamics within a medium of a complex structure.
Time-resolved spectroscopy reveals pronounced variability in three highly ionized absorbers, with Fe~\textsc{xxv}--Fe~\textsc{xxvi} features that appear and disappear on timescales of tens of kiloseconds. This behavior likely reflects a combination of geometrical transits of clumpy gas and ionization-state changes driven by continuum variability. An additional temporary absorption feature in the red wing of the Fe~K$\alpha$ line, consistent with Fe~\textsc{xxv} absorption, indicates a possible transient ultra-fast inflow at $\sim$15\,000~km~s$^{-1}$ ($\sim$5\%~$c$). Finally, the continuum light curve exhibits a tentative $\sim$40~ks oscillatory pattern, accompanied by correlated shifts of a weak, narrow \feka{} emission feature, suggesting dynamic coupling between the continuum and the line-emitting region. Together, these results reveal that the nuclear environment of NGC~3516 is dominated by rapidly evolving, multi-phase gas flows, where accretion, ejection, and ionization processes are tightly coupled on sub-parsec scales.

\end{abstract}


\section{Introduction}\label{sec:intro}

\noindent Rapid accretion of matter onto supermassive black holes (SMBHs) in the centers of galaxies gives rise to the phenomena collectively known as active galactic nuclei (AGN). Their complex central environments, shaped by the accretion and ejection processes, imprint characteristic spectral features across the X-ray band. The Fe~K band (6--8 keV) in particular serves as a uniquely powerful diagnostic of this environment, containing both fluorescent emission lines from the accretion disc and the surrounding structures \citep[e.g.][]{Gallo2023} and absorption features from a range of highly ionized species, most notably Fe~\textsc{xx}--\textsc{xxvi} \citep[e.g.][]{Cappi2006, Tombesi2010}. These spectral features trace gas exposed to extreme gravitational, magnetic, and radiative fields, providing a window into the structure, dynamics, and physical conditions of the immediate vicinity of the SMBH.

Ionized outflows are a ubiquitous and energetically significant component of AGN \citep[e.g.][]{King2015}. They are observed over a broad range of ionization states and outflow velocities, with the most highly ionized ones via spectral signatures in the Fe~K band \citep[e.g.][]{Tombesi2013}. However, the physical origin and launching mechanisms of these outflows remain a matter of active debate. Models invoking radiation pressure on spectral lines, thermal driving, and magnetocentrifugal acceleration have all been proposed \citep[e.g.,][]{Proga2000, Everett2007, Fukumura2010}, yet distinguishing among them requires precise measurements of the ionization structure, kinematics, and—critically—their variability.
The detection and characterization of these features provide crucial insight into the dynamics and physical conditions of gas that links the SMBH to its galactic environment.

Over the past decades, X-ray observations have revealed that AGN environments are not only complex but also highly variable, with substantial changes occurring on timescales from decades to hours or even less \citep[e.g.][]{Gonzalez-Martin2012, Fabian2013, Kara2025}. 
Extracting this information for individual spectral components, however, remains challenging and requires observations that combine high spectral resolution with sufficient signal-to-noise ratio to track the source behavior on its intrinsic variability timescales.

Recent advances in high-resolution X-ray spectroscopy, particularly with the microcalorimeter instrument \textit{Resolve} \citep{Ishisaki2022, Kelley2025} onboard \textit{XRISM} \citep{XRISM, Tashiro2025}, have opened a new window into the time-dependent behavior of the narrow spectral features in the Fe K band. Its high energy resolution enables disentangling individual components manifested in this spectral region \citep{XRISM2024-N4151, XRISM-PDS456, Mehdipour2025, Noda2025}, as well as their time-dependent nature \citep{Xiang2025}.

NGC 3516 is a nearby Seyfert 1.5 galaxy ($z=0.008836$) that has long been known to host an X-ray bright AGN rich in spectral features, particularly in the Fe~K band. A strong \feka{} emission complex, identified already in data from Ginga \citep{Kolman1993}, was subsequently 
recognized to consist of a narrow core and a broader component \citep{Kriss1996}, associated with a relativistically smeared reflection spectrum \citep{Nandra1999, Markowitz2006}. Variability in the emission line was repeatedly observed, affecting both the overall normalization \citep{Nandra1997} and the profile shape itself. Specifically, narrow, absorption-like features in the red wing were seen with \textit{ASCA}, \textit{XMM-Newton}, and the high-energy transmission grating spectrometer onboard \textit{Chandra} \citep{Nandra1999, Turner2002}. Furthermore, a detailed time-resolved analysis of an \textit{XMM-Newton} observation revealed periodic variability in the \feka{} emission, ascribed to a spot in the accretion disc illuminated by a corotating flare located at only a few gravitational radii away from the central black hole \citep{Iwasawa2004}. However, subsequent observations did not reveal this behavior again, and so its nature was left uncertain.
 
The absorption spectrum blueward of the \feka{} emission peak was shown to be no less interesting. In addition to several outflows of lower ionization affecting primarily the soft X-ray band, which have been detected through the years \citep[e.g.][]{Costantini2000, Turner2005, Mehdipour2010, Mehdipour2022}, highly ionized outflows were observed via their Fe~\textsc{xxv}--\textsc{xxvi} absorption features. Just as the soft X-ray absorption, these outflows were also shown to vary, particularly when comparing their signatures between individual observations \citep[e.g.][]{Turner2008, Holczer2012}.
 
In this work, we present a detailed investigation of the Fe K-band in NGC~3516 using the high-resolution spectra obtained with \textit{XRISM}/\textit{Resolve}. Owing to the high flux state of the source during the observation and the exceptional energy resolution of \textit{Resolve}, this dataset provides the most detailed Fe K-band spectrum of NGC 3516 to date. 
We focus on the time-dependent behavior of multiple ionized absorbers detected in the Fe~K band and explore how their variability reflects changes in the underlying continuum emission. By combining spectral and temporal analyses, we aim to disentangle the physical and geometrical drivers of the observed absorption variability and to constrain the properties of the outflowing material. 

The structure of the paper is as follows. Section \ref{sec:data} describes the observations and data reduction. Section \ref{sec:results} presents the results of the modeling of the time-averaged spectrum as well as the time-resolved spectral analysis. Section \ref{sec:discussion} discusses the physical implications of the detected absorbers, focusing on their ionization balance, thermal stability, and connection to the underlying continuum variability. Finally, in Section \ref{sec:summary}, we summarize our conclusions.

\section{Observations and data reduction} \label{sec:data}

\noindent The \textit{XRISM} data were obtained in a joint campaign with \textit{XMM-Newton} \citep{XMM} and \textit{NuSTAR} \citep{NuSTAR}. The observations with the latter two overlapped with that of \textit{XRISM} (see Table \ref{tab:obs} in the Appendix), considerably broadening the spectral band coverage. For the spectral modeling in this paper, data from both \textit{XMM-Newton} and \textit{NuSTAR} were used to reconstruct the spectral energy distribution of the AGN emission. The following subsections describe these observations and the data reduction procedures applied to obtain the calibrated spectra that were used in the subsequent spectral modeling (Sect. \ref{sec:results}).

\subsection{XRISM}

\noindent NGC~3516 was observed with \textit{XRISM} between October 26 and October 29, 2024, with the gate valve closed and the filter wheel in the open position. The data from both \textit{Resolve} and \textit{Xtend} \citep{Noda2025-Xtend} were processed using the JAXA pre-pipeline version 005\_002.20Jun2024\_Build8.012 and the \textit{XRISM} pipeline version 03.00.013.009, with the most recent calibration database (CALDB) release from August 15, 2024.

After applying the standard good time interval (GTI) filtering, the net exposure time of the \textit{Resolve} dataset was 249 ks, out of a total elapsed time of 313 ks. We retained only high-resolution primary events (Hp events), which reduced the total number of events by less than 7\%. For the spectral analysis focused on the Fe~K band, we employed the “large” response matrix, which includes the full energy redistribution in the response but excludes energy channels outside the nominal response range for this configuration. Pixels 12 (calibration source) and 27 (anomalous gain) were excluded from the analysis.

The \textit{Resolve} spectrum was extracted and modeled in the 2--9 keV energy range using the C-statistic as implemented in SPEX \citep[][v3.08.01]{SPEX, SPEX30801}. The spectrum was optimally binned according to the method of \citet{KaastraBleeker2016} using the SPEX command \texttt{rbin}. 

\subsection{XMM-Newton}
\noindent The \textit{XMM-Newton} observation contemporaneous with that of \textit{XRISM} has ObsID 0953790201 and was conducted with all instruments onboard, i.e. the European Photon Imaging Camera (EPIC), Reflection Grating Spectrometer \citep[\textit{RGS};][]{RGS}, and Optical Monitor Optical Monitor \citep[\textit{OM};][]{OM}, in all six available photometric filters onboard, i.e. \textit{UVW2}, \textit{UVM2}, \textit{UVW1}, \textit{U}, \textit{B}, and \textit{V}.
The total exposure of 32.4~ks was taken in the second half of the \textit{XRISM} observation, starting about 165~ks after the beginning of the \textit{XRISM} exposure. The \textit{EPIC-pn} detector operated in Small Window mode with a thin optical blocking filter.

The data were processed with the \textit{XMM-Newton} Science Analysis System (SAS; version 21.0.0) following standard procedures and with the most recent calibration files from October 
2024. Since the observation was affected by background flaring, intervals of high particle background were excluded following the standard procedures, yielding net exposures of 6.7~ks for \textit{EPIC-pn} and 22.7~ks and 22.5~ks for \textit{RGS}~1 and \textit{RGS}~2, respectively.

\subsection{NuSTAR}
\noindent A 55~ks \textit{NuSTAR} observation was carried out contemporaneously with the  \textit{XRISM} pointing, starting on 30 Oct 2024. The \textit{NuSTAR} data were reduced following the standard procedure using the \textsc{heasoft} task \textsc{nupipeline} of the \textit{NuSTAR} Data Analysis Software (\textsc{NuSTARdas}, version 2.1.4), using the calibration files released with the CALDB version 20241104. We applied standard screening criteria, where we filtered for the passages through the SAA, setting the mode to ``optimised'' in \textsc{nucalsaa}. For each of the Focal Plane Module (\textit{FPMA} and \textit{FPMB}), we extracted the source spectra from a circular region with a radius of $50''$, while the background spectra were extracted from two circular regions with a $50''$ radius located on the same detector.  
Light curves in different energy bands were extracted from the same regions using the \textsc{nuproducts} task. After checking for consistency, we combined the spectra and responses from the individual \textit{FPMA} and \textit{FPMB} detectors into a single spectrum. 

\section{Spectral modeling and results}\label{sec:results}

\noindent The spectral analysis was conducted with the cosmological redshift of the source fixed to $z = 0.008836$ \citep{Keel1996}, which corresponds to a luminosity distance of 38.1 Mpc for a flat $\Lambda$CDM cosmology with $H_0 = 70~\mathrm{km~s^{-1}~Mpc^{-1}}$, $\Omega_{\mathrm{m}} = 0.3$, and $\Omega_{\Lambda} = 0.7$. We adopted a black hole mass of $3\times10^{7}~M_\odot$ from the reverberation mapping measurement of \citet{Denney2010}. Absorption by cold Galactic gas was included with the hydrogen column density fixed to $N_{\mathrm{H}} = 4.04\times10^{20}~\mathrm{cm^{-2}}$ \citep{Willingale2013}, modeled with the \texttt{hot} component in SPEX \citep{dePlaa2004, Steenbrugge2005} with the temperature fixed to 0.008 eV. The elemental abundances were fixed to the proto-solar values of \citet{Lodders2009} with the exception of the low-ionization absorbers in the \textit{XMM-Newton}/\textit{RGS} data (see Sect. \ref{ssec:SED}). To account for the attenuation by Galactic dust extinction in the UV--optical band, the model spectrum was reddened with a color excess of $E(B-V) = 0.037$, as measured by \citep{Schlafly2011}, assuming the extinction law of \citet{Cardelli1989} with $R_V = 3.1$.

\subsection{Broad-band Continuum}\label{ssec:SED}

\begin{figure}[t!]
\centering
\includegraphics[width=0.48\textwidth]{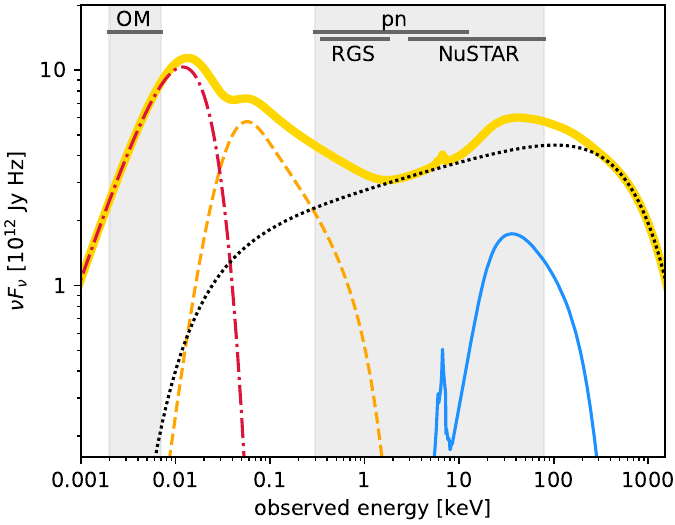}
\caption{SED used for photoionization modeling of the absorbers in NGC~3516, derived with the simultaneous \textit{XMM-Newton} observation. The total SED (in yellow) is a sum of the individual emission components, namely a disk black body (\texttt{dbb}, dashed-dotted red line), warm Comptonization (\texttt{comt}, dashed orange line), exponentially cut-off power law (\texttt{pow}, dotted black line), and relativistically-broadened reflection (\texttt{pexmon*spei}, blue). The spectral bands covered by the campaign are highlighted in grey.
\label{fig:SED}}
\end{figure}

\noindent The broad-band coverage of the observing campaign aided the reconstruction of the spectral energy distribution (SED) of the AGN core. For the modeling, we used the \textit{OM} photometric measurements from all six filters available, the \textit{RGS} spectrum to constrain the soft X-ray absorbers, and \textit{EPIC-pn} and \textit{NuSTAR} to describe the broad-band continuum, combining the high energy resolution of the \textit{RGS} spectrum and the large effective area and spectral coverage of the CCD instruments, as detailed below. 

The \textit{RGS} spectrum was modeled in the spectral range of 7--35~\AA{} and binned by a factor of three (i.e. 30~m\AA{} per bin). To ensure that the higher energy resolution of this dataset drives the fit in this energy range, rather than the high count-rate data from \textit{EPIC-pn}, the \textit{pn} 0.3--1.6~keV spectrum was reduced to a single spectral bin. With this approach, the information on the total flux in the majority of the soft band was retained, and the high-energy end of the \textit{RGS} spectrum (up to 1.77~keV) still overlapped with that of \textit{pn}. This resulted in a seamless transition to the band exclusively covered by \textit{pn}, as the cross-calibration uncertainty \citep{XRISM2025-N3783crosscal} across the full range of the overlapping energies could then be accounted for by a single multiplicative factor of $\approx$1.192, applied to the \textit{RGS} spectrum.

The optimally-binned \textit{NuSTAR} spectrum was fitted from 3~keV to 79~keV, above which the background began to dominate. A good match between \textit{pn} and the combined \textit{NuSTAR} spectrum was achieved with another multiplicative factor, specifically $\approx$0.898 applied to the \textit{NuSTAR} dataset.

The broadband model has the best-fitting parameters listed in Table \ref{tab:sed} in the Appendix and was composed as follows. For the low-energy end of the broad-band emission constrained by the \textit{OM} photometric data, we used a template from \citet{Kinney1996} to account for the stellar emission from the central bulge, and a disc black-body component \citep[\texttt{dbb;}][]{Shakura1973} for the emission from the accretion disc assuming the temperature equivalent to $kT = 10~\rm eV$, where $k$ is the Boltzmann constant. 

The broad-band X-ray continuum was described with a power-law model (\texttt{pow} in SPEX) with the photon index $\Gamma = 1.87 \pm 0.01 $, exponentially cut-off at both ends -- at 13.6 eV and 800 keV. The high-energy cutoff was chosen based on the \textit{NuSTAR} spectrum, which did not show any evidence of a drop in intensity at higher energies.
A Comptonization component (\texttt{comt}) was used to describe the X-ray soft excess, assuming the emission originates from the so-called warm corona \citep{Done2012, Kubota2018} with the seed photons coming from the accretion disc. The electron temperature was constrained from the fit to $0.24_{-0.03}^{+0.04}$~keV, and the optical depth of $\tau = 15_{-2}^{+3} $.

The relativistically broadened reflection from the inner accretion disc was modeled using \texttt{pexmon} \citep{Nandra2007}, convolved with the relativistic profile of \citet{Speith1995} (implemented in SPEX as \texttt{spei}). The parameters to which the high-resolution spectrum from \textit{Resolve} was sensitive to, namely the disc inclination, outer radius, and emissivity profile, were fixed to the best-fitting values from the \textit{Resolve} spectral fit as described below. The remaining parameters of this component were fixed to the default values.

An emission line with a Gaussian profile (\texttt{gaus}) and broadening of $\sim\!1300~\rm km~s^{-1}$ was included to model the additional, narrower Fe~K$\alpha$ feature at 6.4~keV.

In addition to Galactic absorption, the X-ray spectrum was attenuated by several distinct components of photoionized gas associated with the AGN. The absorption in the \textit{RGS} spectrum was well described with two \texttt{pion} \citep{Miller2015, Mehdipour2016} components with $\log \xi \approx 1.37 $ and $\log \xi \approx 2.38 $, where $\xi$ is the ionization parameter defined as the ratio of the ionizing luminosity $L_{\rm ion}$ integrated over 1--1000~Ryd, and the product of the gas hydrogen particle density $n$ and its distance from the ionizing source $r$, $\xi = L_{\rm ion}/(nr^2)$. 
The exploration of partial-covering scenarios revealed that full source coverage was preferred by the data for both ionized outflows. Hence, the covering fraction was constrained to unity in the final model.

Earlier studies \citep{Turner2003, Mehdipour2010} as well as a detailed analysis of atomic abundances in NGC~3516 from this campaign (\textit{E. Behar et al., in prep.}) show that the abundances of nitrogen, and to a smaller extent several other elements, in the photoionized gas in NGC~3516 differ from the proto-solar chemical composition \citep{Lodders2009}. For the purpose of constructing the SED, we allowed only the nitrogen abundance to vary. This additional free parameter was tied between the two low-ionization absorbers and yielded this abundance $3.7_{-0.8}^{+0.9}$ times higher relative to solar, similarly to the works referenced above. To examine the robustness of our results to this choice, we repeated the \textit{Resolve} spectral analysis using SEDs reconstructed for several fixed values of the nitrogen abundance. This test showed that while freeing the nitrogen abundance improves the broadband fit, it does not change the SED in a way that significantly affects our conclusions. 

As detailed below, six more photoionized absorption components were identified in the \textit{Resolve} spectrum. These additional absorbers, with minimal contribution of nitrogen lines due to their high ionization, had this abundance fixed to the proto-solar value and other parameters set to values obtained from the best-fitting model of the time-averaged \textit{Resolve} spectrum.

The resulting SED used to derive the ionic concentrations of the photoionized absorbers in NGC~3516 is presented in Fig.~\ref{fig:SED} and consists of the disc black body emission, warm Comptonization, power-law, and the reflection spectrum, assuming that the remaining components do not significantly contribute to the ionizing flux. The bolometric luminosity corresponding to this model is $\log L_{\rm bol} = 44.1~\rm erg~s^{-1} $ and the ionizing luminosity (integrated over the 1--1000~Ryd range) is $\log L_{\rm ion} = 43.8~\rm erg~s^{-1} $.

\subsection{Time-averaged Fe K-band Spectrum}

\noindent This section describes the \textit{XRISM} \textit{Resolve} spectrum integrated over the entire observation. The best-fitting model of this spectrum serves as a baseline for the subsequent time-resolved analysis.

\begin{figure}[t!]
\centering
\includegraphics[width=0.48\textwidth]{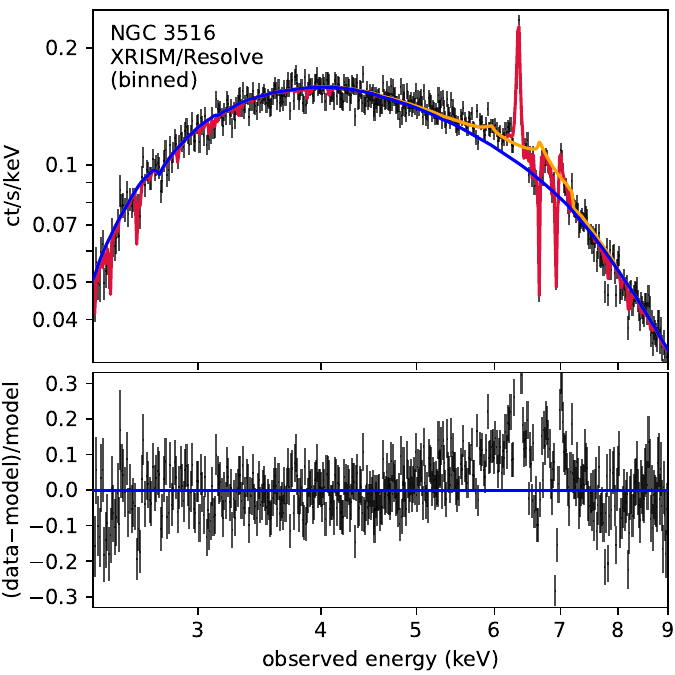}
\caption{Broad-band \textit{XRISM} \textit{Resolve} spectrum of NGC~3516. The red line represents the best-fitting model, the continuum model alone (emission and continuum absorption) is in blue, and a relativistically smeared Fe~K emission is added in orange. The residuals in the bottom panel are plotted with respect to the continuum model to emphasize the excess in emission between 5--8~keV.
\label{fig:bbspecResolve}}
\end{figure}

\subsubsection{Emission Spectrum}\label{sec:em}

\noindent With the ionized absorption being the main objective of this study, we aimed for a simply parametrized, yet accurate description of the underlying emission spectrum. A detailed analysis of the conditions resulting in the Fe K-shell emission lines will be presented in a separate work of \textit{H. Noda et al., in prep.} The modeling adopted here is described below, and the parameters of the emission components constrained from the fit are summarized in Table \ref{tab:emission} in the Appendix.

The continuum in the analyzed spectral range of 2--9~keV was well fitted with a power-law model with the photon index $\Gamma = 1.816_{-0.009}^{+0.008} $ and a small contribution from the Comptonisation component used in the \textit{EPIC-pn} fit to describe the soft excess, with only the normalization left free and the remaining parameters fixed to the XMM-constrained values. Note, however, that the contribution of this component to the \textit{Resolve} spectrum is small and diminishes entirely above 3~keV.

Additional emission, associated with the \feka{} and \fekb{} features, was accounted for with four additional components with distinct line profiles. As a baseline for the emission lines, we adopted the laboratory profiles from \citet{Holzer1997}, with the K$\alpha$ and K$\beta$ complexes scaled such that their ratio was equal to 0.135, assuming an origin in low-ionization gas \citep{Yamaguchi2014}. 

To recover the narrow line cores, no additional broadening was needed, constraining the motion of the emitting material to $ \sigma_v < 75~\rm km~s^{-1}$ (or FWHM~$ < 176~\rm km~s^{-1}$) at 1-$\sigma$ significance. For the remaining line emission components, the \citeauthor{Holzer1997} model was used too, convolved with appropriate profiles as follows. A considerably stronger contribution to the line emission is associated with a broader component with a Gaussian profile (\texttt{gaus}) in our modeling, with the width constrained to $\sigma_v = 1400_{-100}^{+110} ~\rm km~s^{-1}$. A component mimicking the Compton shoulder (\texttt{vcom} profile in SPEX) appearing red-ward of the main \feka{} peak was added as well, although its contribution was rather minimal. 

Finally, a broad and asymmetric excess in emission, possibly indicative of its origin in the inner accretion disc, was modeled with the relativistic \texttt{spei} profile \citep{Speith1995}. Its contribution to the spectrum is visualized in Fig. \ref{fig:bbspecResolve}, in addition to the continuum and the total best-fitting model. The need for enhanced flux in the 6.4--7.2~keV range is apparent at the energies free from line absorption features, and the flux and the width of the \fekb{} emission complex expected from the properties of the \feka{} feature (see the close-up spectrum in Fig. \ref{fig:tavg}). Aided by the unabsorbed maximum blue-ward reach of the emission line at 7.2~keV, the inclination of the inner accretion disc was constrained, within the framework of this model, to $48.8_{-0.2}^{+0.7}~\rm deg$. Additional parameters of this model component that the spectrum was sufficiently sensitive to, namely the emissivity slope and the outer radius, were also determined from the fit, with the remaining fixed to the default values. With this relatively simple model, we obtained a good representation of this broad emission feature and the overall emission spectrum, which allowed us to focus on the narrow absorption features imprinted in it. We stress, however, that more comprehensive models are presently available to study the physical properties of the emission spectrum, which will be addressed in a dedicated study.

As the assumptions made for the emission spectrum modeling could, in principle, affect some of the derived absorption properties, we carried out the spectral analysis also with an alternative model utilizing \texttt{MYTorus} emission components \citep{Yaqoob2012}. This model included additional scattered continuum emission, connected with the same material producing the broad and narrow emission lines. As this modification did not significantly alter the shape of the total emission spectrum, we proceed with the emission model described above. The details of this modeling are given in the Appendix.

\subsubsection{Absorption Spectrum}\label{ssec:abs}

\begin{figure*}[ht!]
\includegraphics[width=1\textwidth]{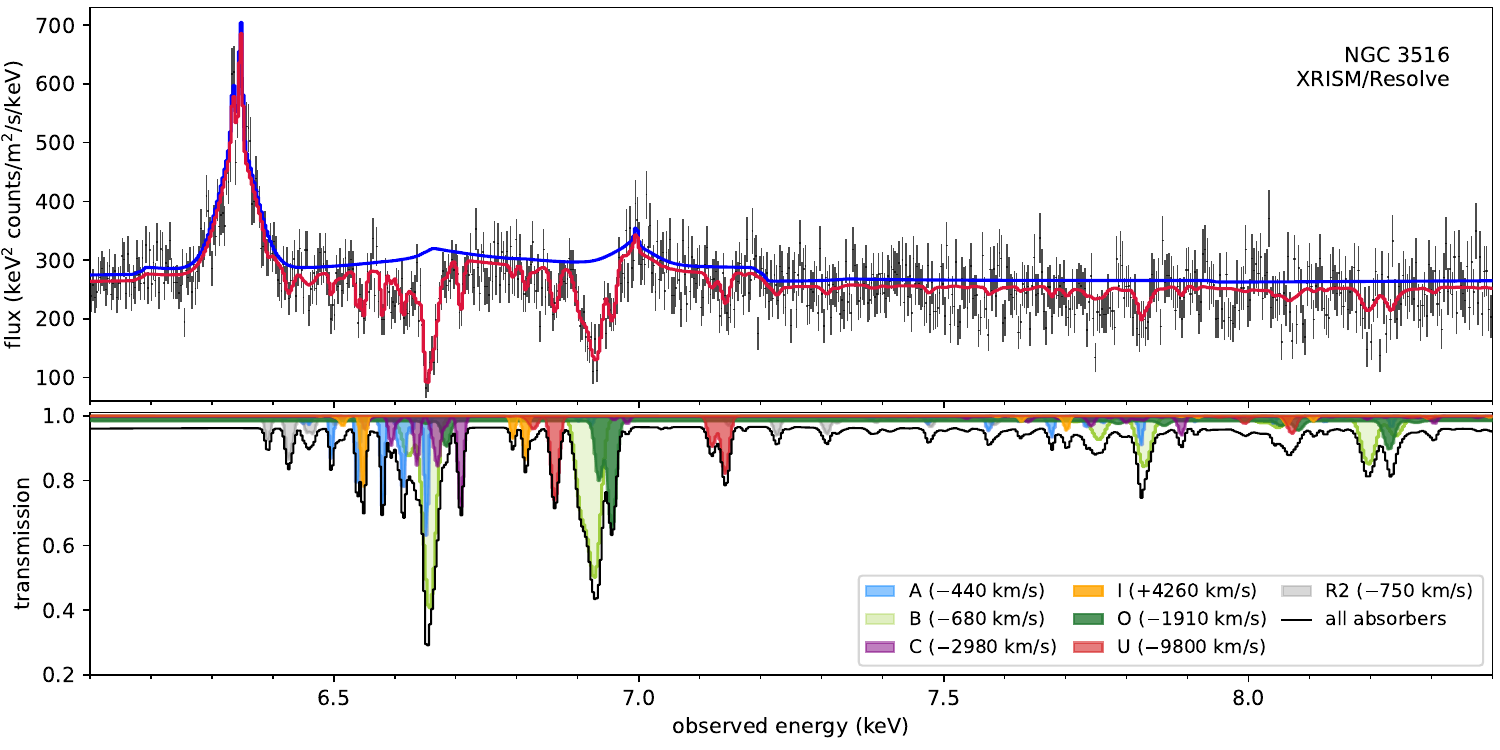}
\caption{The Fe~K band \textit{XRISM} \textit{Resolve} spectrum of NGC~3516 with the best-fitting model. The top panel shows the `optimally' binned data points as used for fitting, overlaid with the best-fitting emission model (blue) and with the absorption included (red). The observed transmission of the individual absorption components at the same energy resolution are plotted in the bottom panel, featuring the absorber constrained from the \textit{RGS} data (R2), persistent absorbers labeled in the order of increasing outflow velocity as A, B, and C, and temporarily appearing absorbers referred to as I (inflow), O (outflow), and U (UFO).
\label{fig:tavg}}
\end{figure*}

\begin{table}[]
    \renewcommand{\arraystretch}{1.2}
    \setlength{\tabcolsep}{3pt}
    \centering
    \begin{tabular}{cccccc}
        \hline\hline
           & $v_{\rm sys}$      &   $\log \xi$          & $\sigma_v$         & $ N_\mathrm{H}$ &$\Delta C$\\
                & $\rm km\,s^{-1}$ & $\rm erg~cm~s^{-1}$ & $\rm km\,s^{-1}$ & $10^{22}~\rm cm^{-2}$\\\hline

A & $-440_{-60}^{+40} $ & $ 3.05_{-0.07}^{+0.03} $ & $ 80_{-80}^{+50} $ & $ 0.63_{-0.09}^{+0.11} $ & 41.8\\
B & $-680_{-60}^{+80} $ & $ 3.53_{-0.07}^{+0.05} $ & $ 370_{-30}^{+50} $ & $ 3.4_{-0.3}^{+0.4} $ & 469.1\\
C & $-2980_{-50}^{+50} $ & $ 3.08_{-0.04}^{+0.04} $ & $ 150_{-70}^{+90} $ & $ 0.41_{-0.11}^{+0.08} $ & 28.2\\
I & $+4260_{-60}^{+80} $ & $ 3.5_{-0.1}^{+0.1} $ & $ 290_{-50}^{+80} $ & $ 1.2_{-0.4}^{+0.2} $ & 12.2\\
O & $-1910_{-90}^{+80} $ & $ 4.0_{-0.2}^{+1.0} $ & $ 300_{-100}^{+100} $ & $ 3.8_{-1.3}^{+0.6} $ & 30.4\\
U & $-9800_{-100}^{+90} $ & $ 3.49_{-0.09}^{+0.08} $ & $ 100_{-100}^{+100} $ & $ 1.4_{-0.4}^{+0.4} $ & 27.2 \\

    \hline
    \end{tabular}
    \caption{\small{Best-fitting parameters of the absorbers constrained from the \textit{XRISM} \textit{Resolve} spectrum integrated over the entire observation. 
    For absorbers A and B, and C, a unity covering fraction $C_\mathrm{f}$ is assumed, but for absorbers I, O, and U, it is set to 0.39, 0.31, and 0.47, respectively, corresponding to the fraction of the net exposure time in which the absorbers are present in the data.}}
    \label{tab:abs-bestfit}
\end{table}

\noindent The 6.5--7.0~keV range is rich in narrow absorption features associated primarily with transitions of Fe~\textsc{xxi}--\textsc{xxvi}, which could be described with $\log \xi $ ranging from 3 to 4 in our modeling. This absorption spectrum was successfully reconstructed with six kinematically distinct photoionized gas components. Continuum and residual line absorption associated with the lower-ionization outflows detected in the \textit{RGS} spectrum was accounted for by two additional photoionized absorption components with parameters fixed to the \textit{RGS} values. All absorbers were modeled with \texttt{xabs} components, which had the ionic column densities obtained with \texttt{pion}, assuming the SED derived from the broad-band fit (Sect. \ref{ssec:SED}). The modeling was performed assuming each of the absorbers was exposed to the unabsorbed broad-band emission spectrum and covered it entirely, with the exception of the narrow emission line cores, which are generally assumed to originate from a more distant material. However, as addressed in detail in the following section, three of these absorbers only appear in a part of the observation. This sporadic nature acts effectively as a reduced source covering fraction, as a significant portion of the underlying emission is unabsorbed when integrating over the entire observation.  

All absorbers show relatively small line widths corresponding to turbulent line broadening of $\lesssim 400~\rm km~s^{-1}$. The largest broadening is exhibited by the strongest absorber with prominent Fe~\textsc{xxv} and Fe~\textsc{xxvi} features. In contrast with this relatively narrow range of line widths, the absorbers span a wide range in systemic velocity, from a mildly relativistic outflow (3~\% of the speed of light) to an inflow at approximately $+4260~\rm km~s^{-1}$. Leveraging the tightly constrained systemic velocity, we hereafter refer to the persistent outflows as A, B, and C, labeled in the order of increasing outflow velocity. The sporadically appearing absorbers are labeled as I (inflow), O (outflow), and U (UFO). The part of the spectrum most affected by the ionized absorption features is displayed in Fig. \ref{fig:tavg} together with the best-fitting model and the ratio of absorbed to unabsorbed model plotted individually for each absorption component. 

As apparent from the plot, this model provides a very good fit to the data, with the achieved $C-$stat of 2437 for 2398 degrees of freedom, lying in the center of the expected range of $2437\pm70$. While a statistically acceptable fit would be achieved with only four absorbers (excluding O and I), the remaining components were required to account for unmodeled absorption lines identified through visual inspection. Additional evidence supporting the identification of these lines arose from the time-resolved spectral analysis described below.

The line broadening was determined from the time-averaged spectrum for the persistent outflows A, B, and C. In the case of the weaker, sporadic absorbers I, O, and U, the broadening was fixed to the best-fitting values determined from a fit of time-resolved spectra described in the following section.
To estimate the uncertainty on the line broadening of these three absorbers, the strongest absorption lines from these components were fitted with a Gaussian line profile, with only the line normalization, centroid energy, and broadening left free. Namely, for I and O, the Fe~\textsc{xxvi} transitions at 6.952~keV and 6.973~keV were used, coupled in energy and broadening, and with the relative normalizations fixed to match the respective oscillator strengths \citep[e.g.][]{Verner1996}. The Fe~\textsc{xxv} line at 6.700~keV was used for U.

The best-fitting values of the systemic velocity, ionization parameter, and the line broadening constrained from this model are given in Table \ref{tab:abs-bestfit}. The total hydrogen column density of the absorbers in the \textit{Resolve} band derived under the assumption of full source coverage is $N_\mathrm{H} = 9.1 \times 10^{22}~\rm cm^{-2}$. However, as mentioned above, this assumption has limited validity due to the sporadically present absorbers, discussed in detail in the following section, and thus this number should be treated as a lower limit.

\subsection{Absorption Variability}\label{ssec:res-varabs}

\noindent Thanks to the high flux state of the source during the observation, \textit{Resolve} registered 0.81 ct~s$^{-1}$ of Hp events in the 2--10~keV band. This high count rate allowed for an analysis of the dataset at a high temporal and spectral resolution. The observation was sliced into 10~ks intervals, and the resulting spectra were examined for intrinsic variability. These snapshot exposures did not allow for spectral modeling at the complexity of the time-integrated spectrum, yet several temporal variations across the observation were identified.

To enable examination of the narrow absorption features free from time-dependent effects of the broadband emission, the continuum variability was accounted for in the following way. Each spectrum was fitted with the best-fitting time-averaged model with only the power-law normalization and photon index allowed to vary. To ensure an accurate description of the continuum shape free from contamination by other potentially variable components, only the energy ranges of 3.5--5~keV and 7.4--9~keV were considered in the fitting. 
Having the continuum profile constrained for each spectrum as a result, the best-fitting power-law component was subtracted from the dataset. 

\begin{figure}[ht!]
\centering
\includegraphics[width=0.46\textwidth]{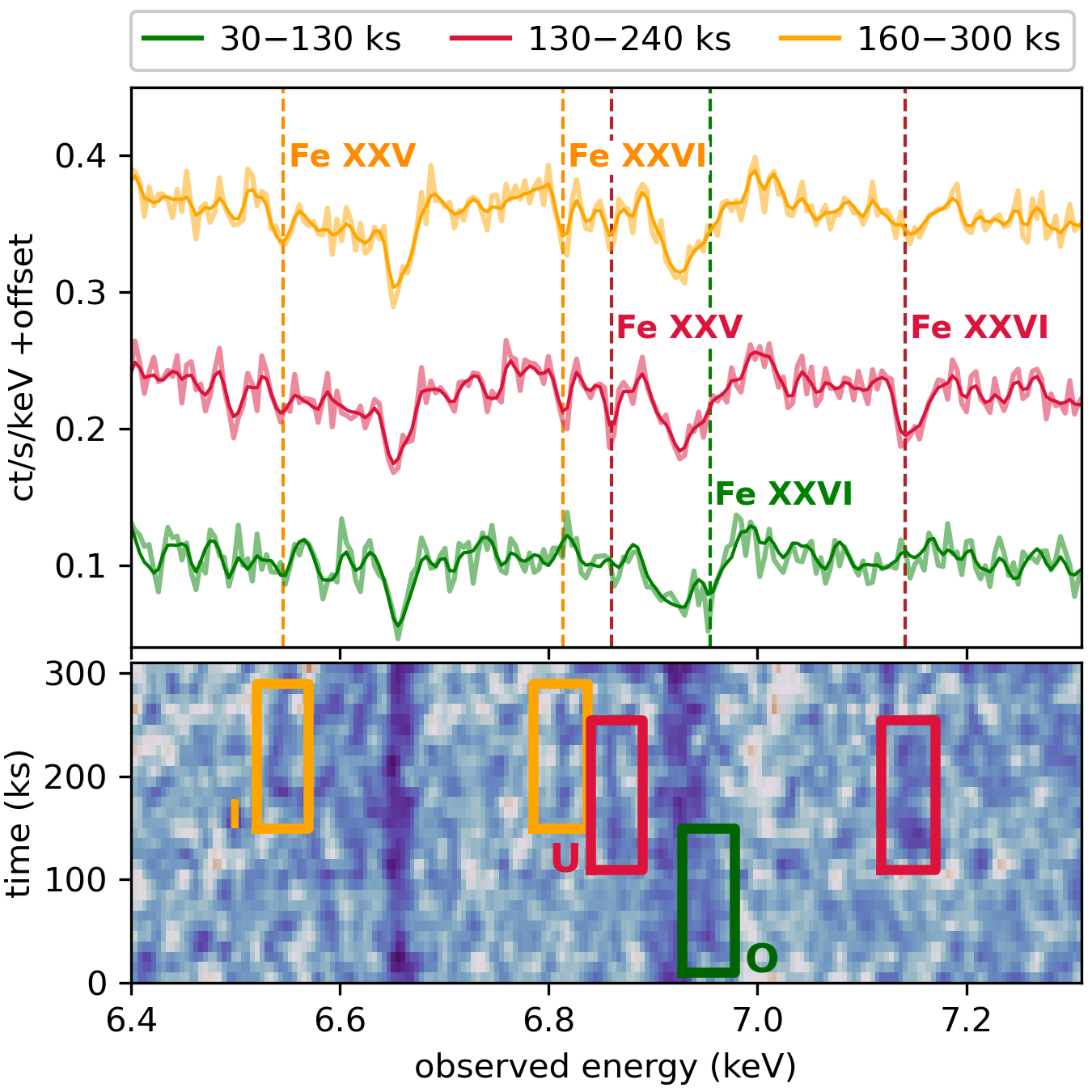}
\caption{Time-dependent nature of the Fe~K-band absorbers. The top panel shows three continuum-subtracted, time-resolved spectra emphasizing the variable absorbers (I, O, U) in the data, with the absorption features marked for each spectrum with dashed lines. The bottom panel presents continuum-subtracted spectra extracted in 10~ks intervals from the start of the observation, shown as a colormap with time increasing along the vertical axis. The positions of variable absorption features are marked with rectangles in the corresponding color. Gaussian smoothing was applied to the map to reduce the effect of noise-related scatter while preserving narrow spectral features in the data. The effect of the smoothing is visible in the top panel, where the observed spectra are given as fainter lines overlaid with their smoothed counterparts. Note that the component U shows temporal overlap with I and O, and so the absorption features `leak' between the displayed spectra.
\label{fig:etdep}}
\end{figure}

Fig.~\ref{fig:etdep} showcases these continuum-corrected spectra in the Fe~K-band and the narrow time-variable features detected in them. In the bottom panel, the snapshot 10~ks spectra are arranged along the vertical axis, visualizing a temporal scan through the observation as a colormap. To reduce the effect of observation noise, the data were convolved with a Gaussian kernel in both dimensions. The effect of this smoothing is apparent from the top panel, where three spectra with reduced exposure are plotted both for the raw data and their smoothed counterpart. The strongest absorption features, well-visible throughout the observation as dark blue stripes at 6.65 and 6.92~keV in the observed frame correspond to the Fe~\textsc{xxv} and Fe~\textsc{xxvi} transitions of absorber B, respectively. 

In addition to these features, several more were detected in the data and identified as sporadically appearing absorbers. The one appearing at the beginning of the observation, noticeable from 30 to approximately 120~ks at 6.96~keV and marked with dark green, is connected to the Fe~\textsc{xxvi} feature of outflow O. At approximately 130~ks from the beginning of the observation, the mildly-relativistic outflow U appears in both Fe~\textsc{xxv} and Fe~\textsc{xxvi} and disappears after about 100~ks, and finally the inflow I appears in the data between approximately 160 and 270 ks in Fe~\textsc{xxv} and Fe~\textsc{xxvi} as well. While the temporary absorbers are relatively weak in the time-averaged spectrum, their concurrent appearance, presence, and disappearance in the data provide not only additional evidence for their existence but also tighter constraints on their physical properties. The spectra displayed in the top panel were extracted from the times of the temporary presence of the absorbers and illustrate the intrinsic spectral variability and the statistical noise of the data with reduced exposure.

The column densities given in Table \ref{tab:abs-bestfit} are given after accounting for the lower effective covering fraction of the sporadic absorbers. The total hydrogen column density of the six absorbers detected with \textit{Resolve} after this correction is $N_\mathrm{H} = 1.9 \times 10^{23}~\rm cm^{-2}$.

These time-resolved spectra were used to determine the line broadening of the temporarily present absorbers. The spectra were modeled with the best-fitting time-averaged model with the parameters of the power law, and those of the absorbers present in a given dataset, left free. The best-fitting values for the line broadening were then adopted for the global spectrum modeling and kept frozen. The cause for the temporary appearance of these absorbers and its implications are discussed in the following section.

\section{Discussion}\label{sec:discussion}

\noindent We have analyzed the \textit{XRISM} observation of the Seyfert 1.5 AGN of NGC~3516 taken in Oct 2024, which yielded a unique dataset rich in spectral features from both emission and absorption sources. Detailed spectral modeling revealed, among others, a reflection spectrum with a prominent broad emission excess, which could be attributed to a relativistically smeared \feka{} line, and six absorbers spanning a wide range of systemic velocities, ranging from an inflow to a mildly relativistic UFO. The observation captured NGC~3516 in a high-flux state, allowing for a time-dependent analysis of the Fe~K band at a high temporal resolution. This analysis resulted in the first detection of intra-observation variability in narrow absorption lines from photoionized material on time-scales of tens of ks seen with \textit{XRISM}/\textit{Resolve}. In this section, we discuss the implications of our findings, aiming to uncover the physical properties of the dynamical environment in NGC~3516 traced by photoionized gas.

Despite the inherent complexity of the Fe~K band, the combination of the high energy resolution of \textit{Resolve} and the high signal-to-noise ratio of the spectrum allowed for the ionized absorption and emission to be separated. Assuming that the broad excess in emission around 6.4~keV can be attributed to relativistic reflection, the apparent cutoff of the blue wing of the relativistic \feka{} line is well detectable in the data and lands at 7.2~keV. This maximal observed energy of the \feka{} photons from the inner accretion disc, blueshifted due to relativistic beaming, is primarily determined by the disc inclination, constrained to $48.8_{-0.2}^{+0.7~}~\rm deg$ from this dataset. In contrast, earlier reports in the literature based on the considerably lower resolution of CCD instruments and various assumptions mention significantly lower values, between 30--40~deg \citep{Wu2001, Nandra2007, Mehdipour2010}. This difference hints at possible past occurrences of ionized absorbers appearing unresolved in the data but effectively suppressing this excess emission. Finally, we caution that the constraint is derived with a rather simple model, and a more detailed analysis is needed to obtain a comprehensive understanding of the emission spectrum.

\subsection{Connection between Absorbers}\label{ssec:thermalstability}

\begin{figure}[t!]
\centering
\includegraphics[width=0.46\textwidth]{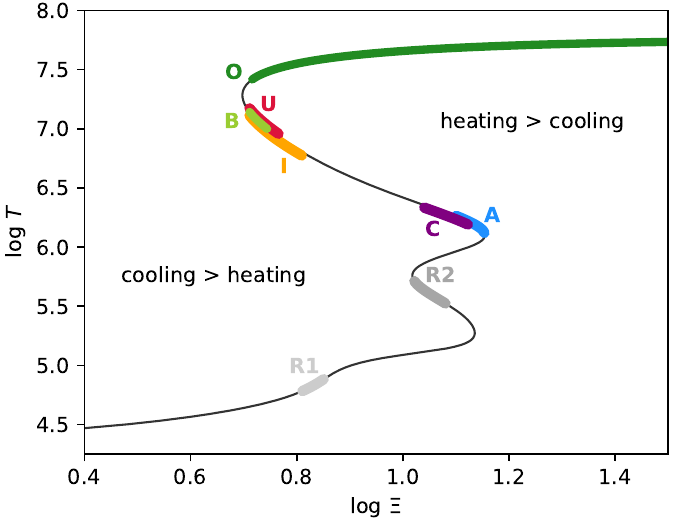}
\caption{Thermal stability curve and the position of the absorbers marked on it, given their ionization state and assuming thermal equilibrium. The size of the absorber markers reflects the 1$\sigma$ uncertainty on $\log \xi$.
\label{fig:s-curve}}
\end{figure}

\noindent The complex absorption spectrum of NGC~3516 revealed eight distinct ionized absorbers -- two low-ionization components and six highly ionized, absorbing in the Fe~K band. 
Even though the absorbers cover a wide range in ionization, column densities, and bulk velocities, they all exhibit a relatively modest turbulent broadening, with velocity dispersions not exceeding approximately $400~\rm km~s^{-1}$, implying that the internal chaotic motions within each absorber are broadly similar.  
A simple picture of an ordered, stratified outflow cannot explain these qualitatively different kinematics. It should also be noted that no significant emission counterparts to the ionized absorbers were detected in the data, which would otherwise result in P-Cygni line profiles. This implies that the observed absorbers do not constitute a part of a large-scale structure, only partially intercepting the line of sight to the radiation source in the AGN core, such as found in the \textit{XRISM}/\textit{Resolve} spectrum of NGC~4151 \citep{Xiang2025}. We note that a possible presence of more complex gas structures is, however, not ruled out by our analysis, and might present an alternative to the scenario derived here \citep{Luminari2024}. A detailed discussion of the distribution of the absorption measure of the gas will be presented in a dedicated study of \textit{E. Behar et al. (in prep.).}

The thermal stability curve of NGC~3516 obtained for the SED described in Sect.~\ref{ssec:SED} does, however, point to a possible connection between the individual absorbers. In Fig. \ref{fig:s-curve}, this curve is displayed as a function of the pressure ratio $\Xi$, defined as the ratio of the radiation pressure and thermal pressure, or $\Xi = \xi/(4\pi c kT)$ \citep{Krolik1981}, where $c$ is the speed of light and $T$ is the electron temperature.  Interestingly, the position of the absorbers in the parts of the curve with a negative gradient reveals that the majority of the absorbers detected in NGC~3516 are thermally unstable, assuming they are in thermal balance with the ionizing radiation. The presence of these unstable regions in NGC~3516 stems from the relatively strong hard-X-ray continuum in this source, which provides a source of additional heating via the Compton process \citep[e.g.][]{KrolikKriss2001, Chakravorty2009}. Although the detailed shape of the stability curve is somewhat sensitive to uncertainties in modeling the broadband SED -- especially in spectral regions where the ionizing continuum is poorly constrained -- the overall presence of unstable zones and the relative positions of the absorbers within them are robust to such variations. This suggests that the different absorber components may be physically connected through dynamical or thermal transitions, rather than representing separate, static layers of a stratified outflow.

The vertical alignment between the low-ionization outflow R1 and the \textit{Resolve}-band absorber B alludes to their possible connection. Detected at a comparable systemic velocity, these two outflows may coexist in thermal equilibrium with each other. In such a scenario, the thermally unstable and highly ionized outflow B would be able to replenish its low-ionization counterpart through runaway cooling. Similarly, such a connection may be in place between the lower-ionization absorber R2 and A, both close to the critical points of the thermal stability curve, taking into account the similarity between their outflow velocities. The case of the sporadic absorbers will be discussed in the following section.

\subsection{Absorption Variability}\label{ssec:disc-absvar}

\noindent Three of the highly-ionized absorbers, I, O, and U, were found to vary in spectral properties during the $\sim$300~ks captured by the \textit{XRISM} observation. This variability was detected via a spectro-temporal scan of the data and manifested as contemporaneous appearance, presence, and disappearance of spectral features (Fe~\textsc{xxv}, Fe~\textsc{xxvi}) associated with the absorbers. This sporadic presence of the absorbers in the data can, in principle, result from two distinct mechanisms. One is due to geometrical effects, i.e., clumps or filaments of gas crossing the line of sight to the compact X-ray source and temporarily obscuring it, and the second is related to changes in the absorber opacity, resulting from changes in ionic concentrations \citep{Nicastro1999}. While the former is dependent solely on the geometrical and kinematic properties of the absorbing material, the latter is induced by changes in the ionizing radiation field, and thus presents an interesting case for exploration of the possible link between the absorbing gas and the radiation responsible for its ionization properties. A connection between the absorption properties and the ionizing SED variability was indeed identified in a recent observation of NGC~3516 by \citet{Mehdipour2022} and could present an explanation for the variability of the absorbers in this dataset.

 In the case of ionization-driven variability, changes in the illuminating X-ray continuum alter the balance between photoionization and recombination processes within the gas, thereby modifying its ionic composition on observable timescales \citep[e.g.][]{Krongold2007, Kaastra2012, Gu2023}. An increase in the ionizing flux can overionize the material, suppressing the line opacity of lower charge states, while simultaneously enhancing the population of more stripped ions. Conversely, a decrease in the flux allows recombination to dominate, repopulating lower charge states and leading to the reappearance of absorption features that may have temporarily vanished. The speed of these transitions depends critically on the gas density, which sets the recombination timescale and thus provides constraints on the location of the absorbers relative to the central engine. Detecting such ionization-induced changes in real time, therefore, not only confirms the physical connection between the absorbers and the central radiation field but also offers a direct diagnostic of the density and radial distance of the absorbing gas. 

Disentangling these two mechanisms requires careful consideration of their distinct observational signatures. Geometrical variability is expected to produce changes in absorption features without a systematic dependence on the continuum flux. Such events may also imprint partial covering signatures, such as simultaneous absorption and transmission of continuum emission. By contrast, ionization-driven variability should track changes in the incident radiation field, producing flux-correlated variations in the relative strengths of ionic species. 

\begin{figure}[ht!]
\includegraphics[width=0.46\textwidth]{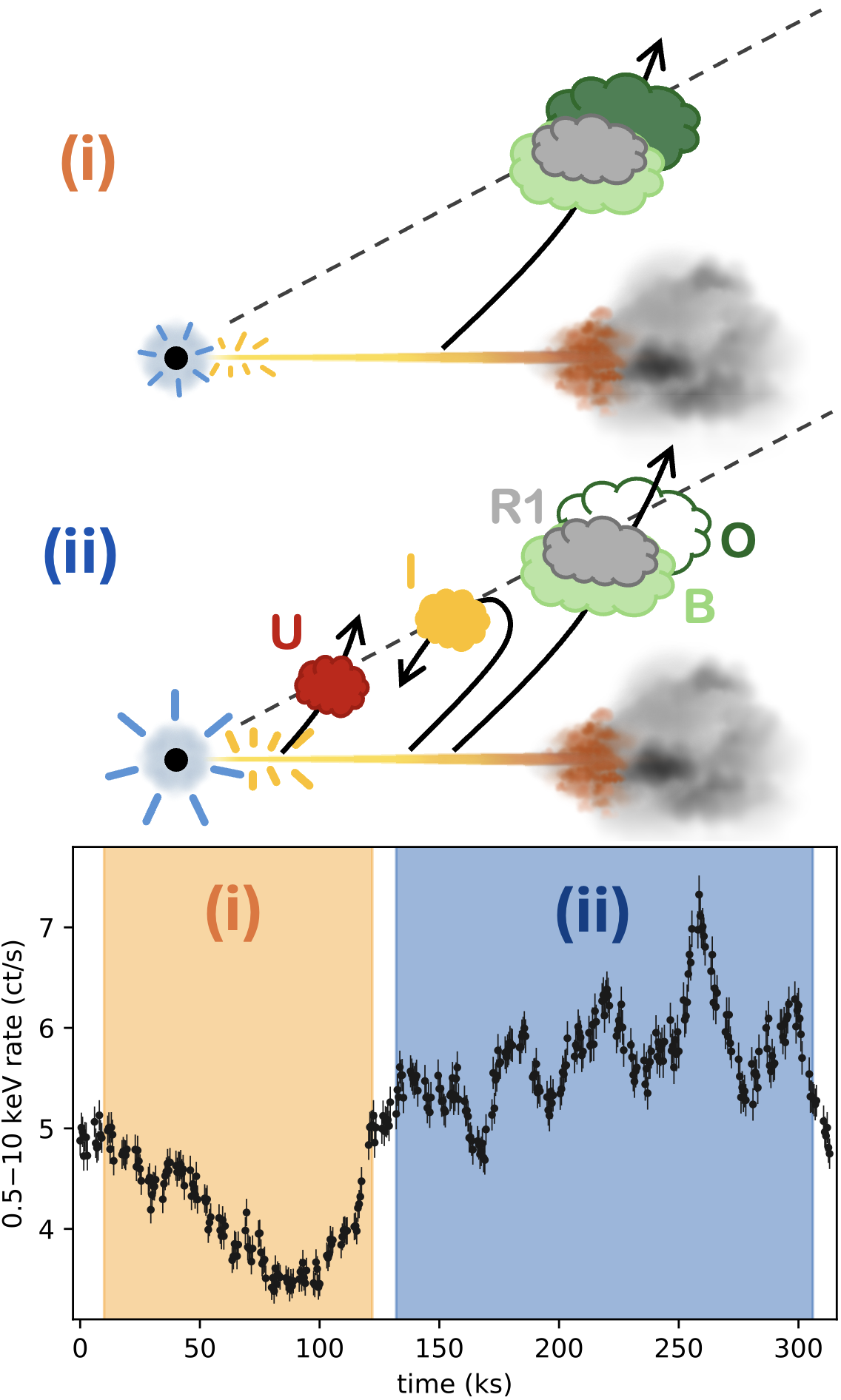}
\caption{Schematic illustration of the X-ray absorbers in NGC~3516 with a focus on the sporadically appearing ones. The bottom panel displays the broadband \textit{XRISM}/\textit{Xtend} 0.5--10~keV light curve binned to 500~s. The initial part with a flux drop (i) and the following high-flux period (ii) are marked with shaded areas, and the corresponding scenarios providing a possible interpretation for the absorption properties are illustrated in the above schematics. In (i), the ionizing continuum is relatively weak and intercepted by the three clouds (representing outflows R1, B, and O). In (ii), the ionizing flux is higher, resulting in overionization of O, which responds to the continuum variability at short timescales. Hence, the absorber is no longer visible in the spectrum despite the overionized material not leaving the line of sight. Additionally, the line of sight is temporarily intercepted by the other two variable absorbers I (inflow) and U (mild-UFO), whose appearance is likely due to geometrical effects rather than a delayed response to the source flux drop. Note that while the absorbers are likely contained within the distance to the broad-line region depicted in orange (inward of the dusty torus in gray), the relative positions of the absorbers are unconstrained and thus are only illustrative. 
\label{fig:cartoon}}
\end{figure}

During the \textit{XRISM} observation, the continuum emission underwent changes that can, indeed, result in changes in the spectral properties of photoionized gas. Specifically, the continuum flux dropped during the first hundred ks, and recovered to the original level at approximately 130~ks from the observation start (Fig. \ref{fig:cartoon}). This decrease affected both the soft excess and the power law, with a 30~\% and 15~\% flux decrease in the 0.5--1~keV and 2--10~keV bands, respectively. As discussed in Sect. \ref{ssec:res-varabs}, the outflow O appears in the data during this drop through the Fe~\textsc{xxvi} absorption feature observed at 6.95~keV and diminishes as the continuum recovers. With a conservative estimate of the low-state ionizing luminosity (integrated over the 1--1000~Ryd range) being 15~\% lower during this period, it is possible that the appearance of the absorption feature was caused by recombination from completely ionized gas. Despite the relatively large uncertainty of the ionization parameter of this absorber, the resulting change in the equivalent width of this Fe~\textsc{xxvi} feature of $\gtrsim$15~\% could explain the observed variability of this absorption component. 
With this assumption, the short response time provides constraints on the gas density. A conservative estimate of the recombination timescale being smaller than 30~ks places a lower limit on the gas hydrogen particle density of $10^6~\rm cm^{-3}$ \citep{Rogantini2022}, which, through the definition of $\xi $, constrains also the gas distance from the ionizing source to $<$$3\times 10^{16}~\rm cm$ or $<$11.6~light days. This would place the outflow within the broad-line region responsible for the broad component of the Fe~K emission lines \citep{Noda2023}, the broadening of which ($\sim$1400~km~s$^{-1}$) corresponds to Keplerian rotation at $10^{17.1}~\rm cm$, assuming the inclination of the inner accretion disc of 48.9~deg, as derived in Sect. \ref{sec:em}. That, in turn, lies within the dust sublimation radius, which can be estimated from the AGN luminosity and the sublimation temperature of silicate or graphite grains to $10^{17.2}\!-\!10^{17.6}~\rm cm$ \citep{MorNetzer2012}. A detailed analysis of the broad component of the \feka{} emission line and its origin is a focus of the work of \textit{H. Noda et al., in prep.}

In the context of the other ionized absorbers detected in NGC~3516, this gas may form a somewhat faster tail of a lower-ionization state, connected with the thermally unstable outflow B and the (stable) outflow R1 (Sect. \ref{ssec:thermalstability}).

In the case of the inflow I and the mildly-relativistic outflow U, which appear in the data with a delay of about 100~ks with respect to the continuum flux drop, the absorption is manifested by features from Fe~\textsc{xxv} and Fe~\textsc{xxvi}. This is caused by an overall lower ionization relative to the outflow O (Table \ref{tab:abs-bestfit}). While the length of the presence of these two absorbers in the data is comparable to the duration of the emission flux drop, which may suggest a link between the two, the magnitude of the delay and the concurrent presence of the absorption lines from both species make the scenario of the changing ionic composition less likely to explain the observed data alone. The delayed response to the ionizing continuum is a result of the density-dependent recombination timescale. In contrast, the ionization timescale is density independent. As a result, the changes in ionic concentrations are smoother and less prominent for slowly responding media than for a rapidly responding gas. Additionally, the change in the ionization state of the gas that would be needed for both Fe~\textsc{xxv} and Fe~\textsc{xxvi} to appear/disappear is considerably larger than what the X-ray continuum behavior suggests. With these considerations, we conclude that geometrical effects likely play a dominant role in the detected variability of these two absorbers. 

Assuming the temporary appearance is due to their transverse motion across the line of sight, broad constraints on the location of these absorbing clouds can be placed. The crossing time of $\sim$100~ks corresponds to a transverse velocity of $\sim$$4\times 10^3~ \rm km~s^{-1}$ for a clump of gas crossing the X-ray source with a diameter of $\sim$10~gravitational radii of the central supermassive black hole with the mass of $3\times10^{7}~M_\odot$. If associated with Keplerian motion, the gas would have a hydrogen particle density of $\sim$$10^8~\rm cm^{-3}$ and would be located at $ 10^{16}\!-\!10^{16.5}~ \rm cm$, consistent with the broad-line region \citep[e.g.][]{Noda2023}. Interestingly, reverberation mapping of the strongly asymmetric H\,$\beta$ line observed in NGC~3516 revealed an inflow component in the (H\,$\beta$-emitting) broad-line region gas at a radial velocity of $\sim\!6400~ \rm km~s^{-1}$ \citep{Denney2009, Oknyansky2021}, not dissimilar to the velocity observed in the inflow I reported here ($\sim\!4300~ \rm km~s^{-1}$).

A schematic illustration of the X-ray absorbers in NGC~3516, with particular emphasis on those appearing sporadically during the \textit{XRISM} observation, is presented in Fig.~\ref{fig:cartoon}. For clarity, and given the lack of constraints on their geometrical properties, we omit the absorbers A, C, and R2 in this cartoon. The lower panel of the figure shows the broadband \textit{XRISM}/\textit{Xtend} light curve extracted from the 0.5–10~keV energy range. Two characteristic phases of the varying continuum flux are highlighted: an initial low-flux episode (i) and a subsequent high-flux period (ii). A scenario proposed to explain the observed spectral properties during these phases is depicted in the two illustrations above.

During phase (i), the continuum emission is relatively weak, and the line of sight intersects three absorbing structures, associated with outflows R1, B, and O. Under these conditions, the ionization balance within absorber O remains such that transitions of Fe~\textsc{xxvi} imprint a detectable absorption feature on the spectrum. In contrast, during phase (ii), the continuum flux increases significantly, and the stronger ionizing radiation alters the ionization balance of the same absorber. The enhanced ionization state drives O into an overionized regime where the aforementioned ionic species becomes depopulated, thereby suppressing its spectral signatures. Importantly, this disappearance does not require the absorber to move out of the line of sight physically; rather, it reflects the sensitivity of its opacity to changes in the illuminating radiation field, indicating a close causal link between the continuum variability and absorber properties on short timescales.

In addition to the ionization-driven disappearance of O, the high-flux phase (ii) is also characterized by the temporary appearance of two further absorbers, labeled I (an inflowing component) and U (a mildly relativistic outflow or UFO). Their emergence is unlikely to result from ionization changes, as their variability does not directly correlate with the observed flux trend. Instead, their presence is more plausibly attributed to geometrical effects, such as discrete clumps of gas crossing the line of sight and intermittently obscuring the central X-ray source. This combination of ionization-driven and geometrical variability highlights the complex and dynamic nature of the circumnuclear environment in NGC~3516, where multiple absorbers with distinct physical origins contribute to the observed spectrum. 

It should be emphasized that the spatial arrangement of the absorbers in Fig.~\ref{fig:cartoon} is illustrative only, as their exact radial positions and relative locations along the line of sight are unconstrained by the present data. Nonetheless, the schematic serves to demonstrate how different mechanisms—rapid ionization responses and line-of-sight occultations—can operate simultaneously and shape the observed absorption variability in this source.

Finally, it should be noted that the relatively hard SED to which the outflows in NGC~3516 are exposed is unlikely to be the dominant cause of their outward acceleration. While radiation pressure can, in principle, accelerate AGN winds to velocities exceeding a thousand km~s$^{-1}$ \citep[e.g.][]{Mushotzky1972, Scargle1973, Proga2000}, the large fraction of hard X-ray photons relative to the UV flux in the reconstructed SED would lead to rapid overionization and a consequential decrease in momentum transfer. Therefore, it is reasonable to expect that magnetic pressure or magnetocentrifugal forces dominate the acceleration \citep[e.g.][]{Fukumura2010}. Overall, the combination of both inflowing and outflowing material implies a dynamic environment in which magnetic and radiative forces compete, potentially resulting in cyclical episodes of launching and fallback. 

\begin{figure}[t!]
\includegraphics[width=0.46\textwidth]{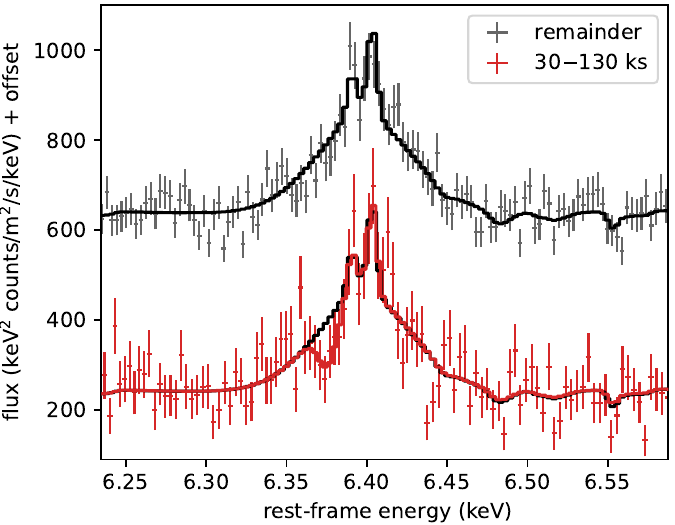}
\caption{Variable absorption feature in the red wing of the broad \feka{} line. The red data points correspond to the spectrum extracted from the 30--130~ks from the beginning of the observation, and the black data points (with a vertical offset) the spectrum from the remaining part of the observation. The solid lines represent the best-fitting models assuming the time-averaged parameters, with the exception of the power-law normalization and photon index. For the red dataset, an absorption line with a Gaussian profile was added with the line normalization, centroid energy, and width left free. 
\label{fig:ufi}}
\end{figure}

\subsection{Variability in the \feka{} Emission Line Profile}

\noindent The Gaussian profile and the additional narrow line core provide a satisfactory description of the prominent emission line profile. However, variability in the line profile was, nevertheless, detected in the time-resolved analysis.

\subsubsection{Ultra-fast Inflow}

\noindent During the observation, a narrow, absorption-like feature temporarily appeared in the red wing of the \feka{} emission line. This feature is well visible in Fig.~\ref{fig:ufi}, where the \textit{Resolve} spectrum from 30--130~ks from the observation start was extracted. The high-time-resolution scan through the \textit{Resolve} observation detailed in Sect. \ref{ssec:res-varabs} further revealed that this feature was persistently present in at least six contiguous 10~ks spectral slices between approximately 40--100~ks from the observation start. This behavior is analogous to the temporary absorbers discussed in the previous sections, albeit detected only as a single spectral feature. By contrast, no other remaining line-like features exhibit comparable persistence across adjacent time slices. The absence of similar temporal coherence in the remaining residuals indicates that they are consistent with statistical fluctuations, reinforcing the interpretation of the 6.37~keV feature as real.

When modeled with a Gaussian profile, the line is detected at 3$\sigma$ significance with the rest-frame centroid energy constrained to $6.37\pm 0.03~\rm keV$ and the width to $\sigma_{v} = 250_{-120}^{+90}~\rm km~s^{-1}$. This feature does not seem to have a similarly strong counterpart that would aid unambiguous identification of its origin. This suggests the absorption may originate from Fe~\textsc{xxv}, with a relatively large redshift with respect to the NGC~3516 rest frame. Indeed, using a photoionized gas model instead of a single Gaussian line allows the feature to be modeled as such, with $\log \xi = 3.3_{-0.3}^{+0.2} $, $N_\mathrm{H} = 3_{-2}^{+3}\times 10^{21}~\rm cm^{-2}$ and the systemic velocity  $v_\mathrm{sys} = 14\,800_{-400}^{+200}~\rm km~s^{-1} $. We note that thanks to the high spectral resolution of \textit{Resolve}, the interpretation of the feature as absorption from Fe~\textsc{xxvi} doublet at 6.952 and 6.973~keV with oscillator strengths roughly in a ratio 1:2 \citep[e.g.][]{Verner1996}, is disfavored by the data, given its narrow and symmetric appearance in the spectrum.

Such a large redshift, corresponding to an inflow at 5~\% of the speed of light relative to the source rest frame, places this additional absorber among ultra-fast inflows, which have been rarely detected thus far in AGN spectra. Yet, narrow absorption features in the red wing of \feka{} were reported in NGC~3516 before. While at a much lower spectral resolution, relatively narrow modulation of the red wing was observed in \textit{ASCA} data \citep{Nandra1999}, and subsequently also in simultaneous observation with \textit{XMM-Newton} \textit{pn} and \textit{Chandra HETGS} \citep{Turner2002}. Perhaps the best example of highly ionized inflows in the literature is of those observed in a luminous Seyfert 1 AGN of PG1211+143, detected through highly redshifted lines identified as Fe~\textsc{xxvi} with \textit{Chandra LETGS} \citet{Reeves2005}. While relying on a lower-resolution CCD spectrum of \textit{XMM-Newton} \textit{pn}, \citet{Pounds2018}  detected a highly redshifted absorber in this AGN in resonance lines of several species, thanks to which it was unambiguously identified in the data. This inflow was also of a temporary nature, as it was not detectable in another observation separated by only two weeks. The inflow was later detected also at lower energies in stacked \textit{RGS} spectra from several orbits \citep{pounds2024}. Finally, a similar complexity of the \feka{} red wing was recently found in the \textit{Resolve} spectrum of Mrk~279 \citep{Miller2025}. However, in that case, the source flux did not allow for the temporal aspect to be tested.

Despite the large inferred inflow velocity, the absorption feature remains intrinsically narrow, which is difficult to reconcile with a scenario in which gas is freely falling from large radii. In such a case, a substantial velocity gradient along the line of sight would be expected, producing broader absorption than is observed. The persistence of a narrow profile instead hints that the inflowing gas may be concentrated in a compact structure rather than distributed smoothly. This raises the possibility that the material is confined or guided by an external agent, such as magnetic fields, allowing it to maintain its integrity as it moves inward. If so, the detection of such a structure along our line of sight suggests that similar confinement mechanisms may operate on both inflowing and outflowing gas, pointing toward a more dynamic and interconnected medium than a simple picture of separate, independent inflows and winds.

A rather subtle inflow such as the one identified here would be undetectable with the instruments available thus far, which suggests that direct inflow of gas towards the supermassive black hole may be more prevalent than generally assumed. This further emphasizes the value of high-throughput, high-resolution microcalorimeter spectra for the understanding of black hole accretion.

\subsubsection{Short Time-scale Variability}

\begin{figure}[t!]
\includegraphics[width=0.46\textwidth]{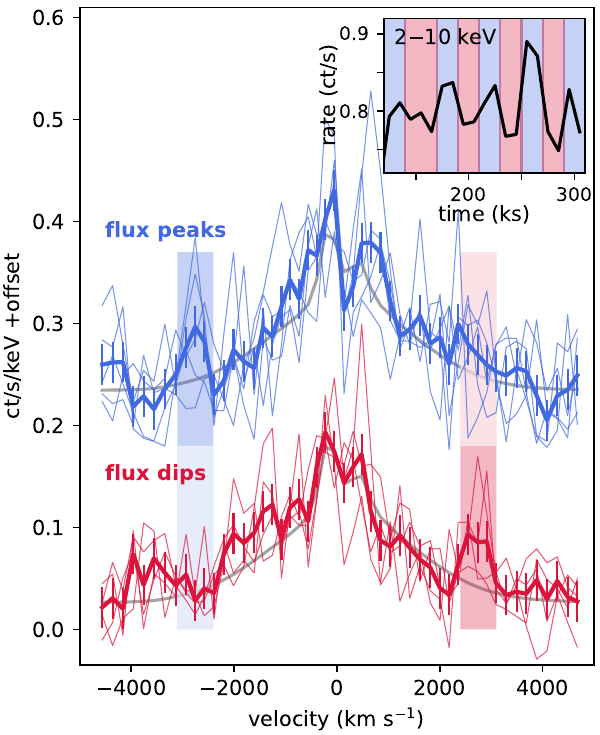}
\caption{\feka{} profile of NGC~3516 observed with \textit{XRISM}/\textit{Resolve}, displayed as a function of velocity centered at 6.4 keV. The main panel showcases the side peak appearing at a shift of $\approx 2700\rm ~km~s^{-1}$ blueward and redward of the line center as the continuum flux peaks (plotted in blue with a vertical offset) and dips (red lines), respectively. The thicker line represents the mean profile, and the thin lines show the spectra from the individual slices of the observation. The spectra were extracted with a 20~ks timestep, starting from the onset of the apparent periodic fluctuations of the continuum emission. The gray line represents the best-fitting model of the time-averaged spectrum scaled in normalization to match the continuum flux of the red and blue spectra.
{\it Inset}: 2--10~keV \textit{XRISM}/\textit{Xtend} light curve with the flux peaks and dips highlighted in blue and red, respectively.
\label{fig:rockingFeKa}}
\end{figure}

\noindent The continuum light curve of the final 190~ks of the observation exhibits an oscillatory pattern, with a tentative characteristic period of approximately 40~ks. The limited number of observed cycles (4--5) prevents a statistically significant confirmation of this periodicity, and so the observed behavior remains consistent with the stochastic variability typically seen in AGN X-ray light curves. Yet, the apparent regularity of the oscillations is intriguing and may hint at the presence of a quasi-periodic process operating in the vicinity of the black hole, such as modulation by disc inhomogeneities, orbiting hot spots, or oscillatory modes in the corona or inner disc region. 

Notably, these flux variations appear to be imprinted onto the shape of the broad Fe~K$\alpha$ emission line (Fig.~\ref{fig:rockingFeKa}), suggesting that the continuum and line-emitting regions are dynamically coupled. In time-resolved spectra extracted around individual flux minima and maxima, a transient narrow emission feature is observed to shift alternately to the red and blue sides of the central 6.4~keV line centroid, corresponding to a velocity offset of approximately $\pm$2700~km~s$^{-1}$. The feature exhibits a flux of $\lesssim\! 3 \times 10^{-13}~\rm erg~s^{-1}~cm^{-2}$, contributing less than 5\% to the total line flux. 

We note that this is the second time a rapid, periodic variability is detected in the \feka{} line in NGC~3516. \citet{Iwasawa2004} reported a similar modulation captured in an observation performed with \textit{XMM-Newton}, with a period of 25~ks. This periodic variability, associated with an additional peak to the stable emission profile centered at 6.4~keV in the source rest frame, was moving periodically in energy, similarly to what was observed with \textit{XRISM} in the dataset analyzed in this work. These changes were not observed again, suggesting a short-lived nature of the event. 
Periodic, short-time-scale variability in the broad red wing of the \feka{} emission line correlated with the continuum has, however, also been detected in another Seyfert galaxy. NGC~3783 showed a very similar behavior to that previously observed in NGC~3516, with a period of 27~ks measured in five cycles of an \textit{XMM-Newton} observation \citep{Tombesi2007}.

Having analyzed the entire observation for the presence of the side peaks, we conclude that they emerge concurrently with the onset of the apparent oscillations in the continuum emission. If real, this oscillatory ``rocking'' motion of the narrow component could trace azimuthal motion or localized perturbations within the inner accretion disc, possibly associated with a transient overdensity or a hotspot. With an orbital period of 40~ks, the emitter would be located at a radius of $\sim$14~$r_{\rm g}$. With this orbital period, however, the Doppler shift at the extremal points would be much larger than the observed $\pm$2700~km~s$^{-1}$, unless the orbit is observed at a substantially smaller inclination than the bulk of the accretion disc. Alternatively, the observed variability could arise from changes in the opacity of an ionized medium that partially covers the X-ray source, producing alternating red- and blue-shifted absorption features as the line-of-sight conditions evolve. A further possibility is that the feature originates in emission from compact, rapidly moving clouds located near the inner edge of the optical broad-line region, which may transiently respond to the variable X-ray illumination \citep{Oknyansky2021}. 

Discriminating between these scenarios requires more extensive temporal coverage and higher statistical precision than currently available. However, the detection of such subtle, velocity-resolved line variations demonstrates the remarkable potential of \textit{XRISM}/\textit{Resolve} for time-domain X-ray spectroscopy. Its unprecedented combination of high energy resolution and throughput enables direct tracking of short-timescale spectral changes in the Fe~K band, providing new access to the dynamical behavior of matter in the immediate vicinity of the supermassive black hole. 

\section{Summary}\label{sec:summary}

\noindent We have presented a detailed analysis of the \textit{XRISM} observation of the AGN in the Seyfert~1.5 galaxy NGC~3516, which has brought the first high-resolution X-ray spectroscopic view of this source in the Fe~K band. The exposure spanning $\sim$310~ks of elapsed time revealed an exceptionally rich and time-variable absorption spectrum superimposed on a complex relativistic and distant reflection continuum. The key results can be summarized as follows:

1.~The continuum of NGC~3516 is well described by a power law, modified by both relativistic and distant reflection. The emission excess around \feka{} and \fekb{} complexes can be successfully modeled using a relativistically broadened reflection component, inclined at $48.8_{-0.2}^{+0.7~}~\rm deg$, combined with two symmetric and narrower line emission components with the broadening corresponding to $1400_{-100}^{+110} ~\rm km~s^{-1}$ and $<\!75~\rm km~s^{-1}$.

2.~In addition to two low-ionization outflows with features dominant in the \textit{RGS} band, the \textit{Resolve} spectrum revealed six more distinct absorption components with no detected emission counterparts. These absorbers span an order of magnitude in ionization parameter and a wide range of systemic velocities, from an inflow to a mildly relativistic ultra-fast outflow. The turbulent velocity broadening of all absorption components is relatively small ($\lesssim$$400~\rm km~s^{-1}$), and points to similar internal kinematics across components. While some absorbers share comparable ionization and velocity characteristics, others display markedly different kinematic signatures, indicating that the absorbing medium has a complex structure and is not a single, continuous outflow. Furthermore, the majority of the detected absorbers are likely thermally unstable and thus may represent transient structures.

3.~Time-resolved spectroscopy of the observation reveals significant changes in the absorption features associated with three highly ionized components (labeled I, O, and U). Their spectral signatures, namely Fe~\textsc{xxv} and Fe~\textsc{xxvi} lines, appear and disappear over tens of kiloseconds. Two distinct mechanisms may explain this behavior: (a) geometrical transits of discrete clumps across the line of sight, or (b) ionization-state changes driven by variations in the incident continuum. The observed correlation between the absorber visibility and the continuum level, particularly for component O, favors ionization-driven variability, whereas the intermittent appearance of components I and U is more consistent with geometric effects.

4.~An additional transient absorption-like feature was detected at 3$\sigma$ significance in the red wing of the \feka{} emission line, possibly revealing the presence of a brief, ultra-fast inflow event. The feature, centered at 6.37 keV, corresponds to an inflow velocity of approximately $14\,800~\rm km~s^{-1}$ ($\approx$5~\%~$c$) if attributed to Fe \textsc{xxv} absorption. This narrow and symmetric feature persisted for about 60~ks and was not accompanied by a comparable blueshifted counterpart, suggesting a localized and short-lived episode of gas accretion toward the black hole.

5.~Latter part of the observation revealed oscillatory variability in the continuum light curve, with a tentative period of $\approx$40 ks. These flux oscillations appeared to modulate the \feka{} line profile, where a weak, transient narrow emission feature alternately shifted to red and blue energies by about $\pm$2700~km~s$^{-1}$. This ``rocking'' motion suggests dynamic coupling between the continuum source and the line-emitting region, potentially linked to orbiting hotspots, disc inhomogeneities, or oscillations in the corona. A similar phenomenon was previously reported in NGC~3516 \citep{Iwasawa2004} and NGC~3783 \citep{Tombesi2007}, implying that such rapid variability may trace transient structures in the innermost accretion disc of these objects.

\begin{acknowledgments}
AJ acknowledges support from NASA Grant 80NSSC25K0082. FT acknowledges funding from the European Union -- Next Generation EU, PRIN/MUR 2022 (2022K9N5B4). The Technion team was supported by the Israel Science Foundation (grant No. 2617/25). 
\end{acknowledgments}

\appendix

\section{Emission modeling and additional material}

\begin{figure*}[h]
\includegraphics[width=1\textwidth]{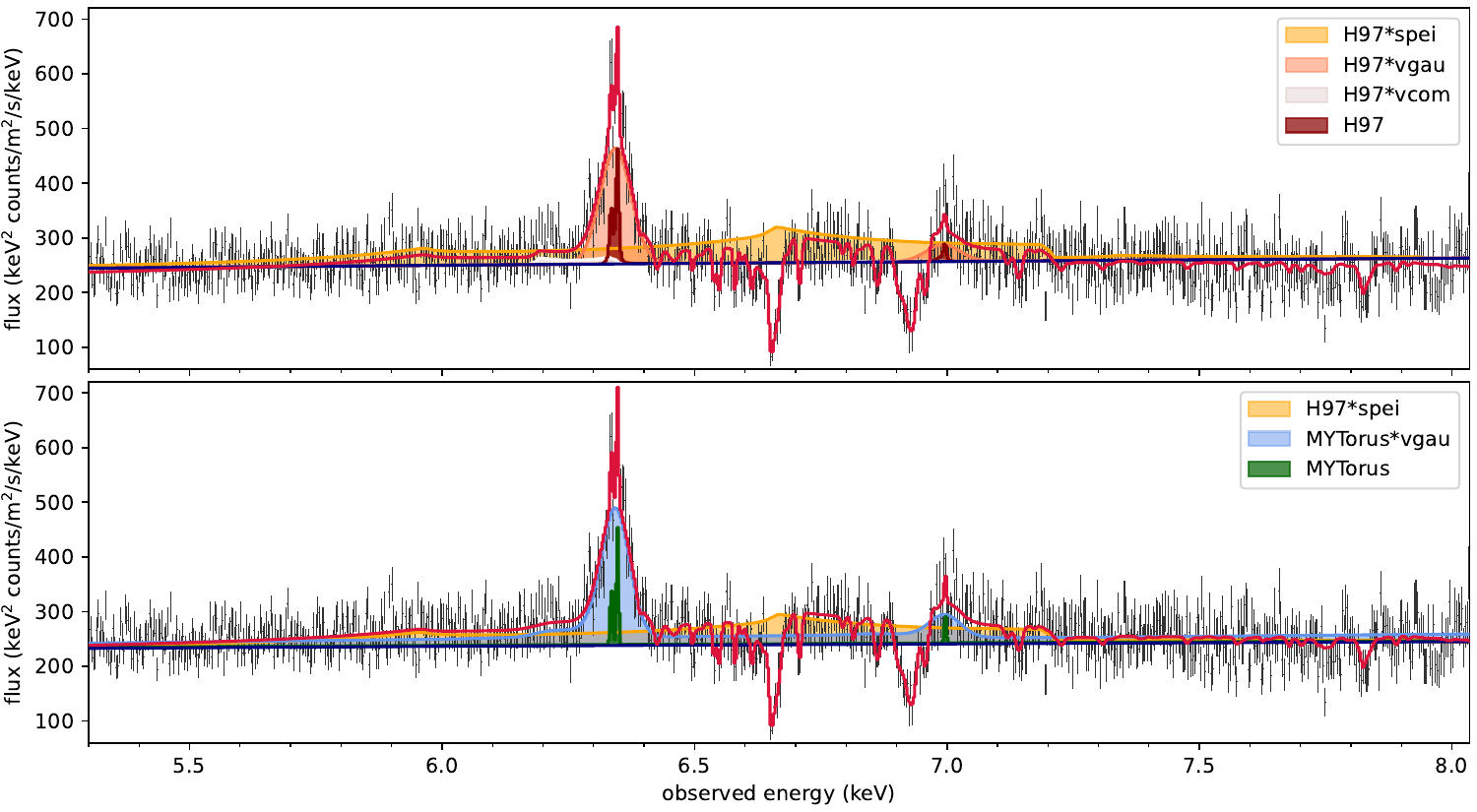}
\caption{Comparison of the adopted time-averaged model (top) of the \textit{XRISM}/\textit{Resolve} spectrum of NGC~3516 with an alternative one (bottom). The top-panel emission spectrum model is composed of the \citet{Holzer1997} profile for \feka{} and \fekb{} (brown), convolved with different broadening models: relativistic (yellow), Gaussian (orange), and Compton shoulder (light pink). In the bottom panel, the \citeauthor{Holzer1997} profile is convolved with the relativistic broadening (orange). The rest of the emission is modeled with \texttt{MYTorus}, which includes scattered emission in addition to the lines, convolved with a Gaussian profile (blue) and with no additional broadening (green). A power-law continuum is used in both models (black). The total best-fitting models, including the absorption components presented in Sect. \ref{ssec:abs}, are depicted in each panel with a red line.}
\label{fig:2em}
\end{figure*}

\noindent As an alternative description of the Fe K–band emission to the model presented in Sect. \ref{sec:em}, we also tested a scenario in which the distant reprocessing component is replaced by the physically motivated \texttt{MYTorus} components \citep{Yaqoob2012}. This model self-consistently accounts for fluorescent line emission and Compton-scattered continuum from a toroidal, neutral or weakly ionized medium. In this configuration (Fig.~\ref{fig:2em}, bottom panel), the broad and asymmetric excess component is still modeled using the \citet{Holzer1997} \feka{} and \fekb{} line profiles convolved with the relativistic blurring kernel \texttt{spei}, but the remaining narrow and intermediate-width emission features are reproduced by two \texttt{MYTorus} components representing distinct reprocessing regions. One \texttt{MYTorus} component is convolved with a Gaussian kernel to mimic moderate velocity broadening, while the other is left unbroadened to represent emission from more distant material that produces the narrow core. 

Importantly, both models provide a very good, statistically indistinguishable fit to the data. Yet, expectably, some of the emission model parameters somewhat vary in their normalizations, most notably the relativistic line, a part of which in the baseline model is replaced by the scatter continuum added through the \texttt{MYTorus} components. Overall, however, the models yield largely consistent parameters of the absorption spectrum components.

Unlike the purely line-emission components adopted in the baseline model, \texttt{MYTorus} naturally includes the associated Compton-scattered continuum and line shoulders. 
Although this alternative model provides a comparably good fit to the \textit{Resolve} spectrum, it introduces additional physical assumptions about the geometry and column density of the reprocessor, and we therefore retain the simpler baseline model for the time-resolved analysis while using \texttt{MYTorus} as a consistency check on the nature of the distant reflection.

\begin{table}[h]
\begin{center}
\begin{tabular}{lccc} 
\hline\hline
Mission & ObsID & Start Time & Exposure (ks) \\
\hline
\textit{XRISM} & 201034010 & 2024-10-26 01:16:43 & 249 \\
\textit{XMM-Newton} & 0953790201 & 2024-10-28 09:00:43 & 32 \\
\textit{NuSTAR} & 91001651002 & 2024-10-28 16:11:08 & 55 \\
\hline
\end{tabular}
\end{center}
\caption{\small{Details of the NGC~3516 observations taken during the joint \textit{XRISM}, \textit{XMM-Newton}, and \textit{NuSTAR} observing campaign analyzed in this work.}}
\label{tab:obs}
\end{table}

{\small
\begin{table}[h]
\hspace{-1.0cm}
\begin{center}
\begin{tabular}{lll}
\hline\hline
{\bf Component} & {\bf Parameter (units)} & {\bf Value} \\
\hline

{\tt pion} (R1) & $ N_{\rm H}~(10^{21}~\rm cm^{-2})$ & $4.3_{-0.3}^{+0.3}$  \\
          & $\log\xi~(\rm erg~cm~s^{-1})$ & $1.37_{-0.07}^{+0.08}$  \\
          & $\sigma_v~(\rm km~s^{-1})$ & $90_{-20}^{+20}$  \\
          & $v_{\rm sys}~(\rm km~s^{-1})$ & $-520_{-60}^{+60}$  \\
          & $A_{\rm N}/A_{\rm N,~solar}$ & $3.7_{-0.8}^{+0.9}$$^*$  \\
{\tt pion} (R2) & $ N_{\rm H}~(10^{21}~\rm cm^{-2})$ & $6.3_{-0.1}^{+0.1}$  \\
          & $\log\xi~(\rm erg~cm~s^{-1})$ & $2.38_{-0.07}^{+0.08}$  \\
          & $\sigma_v~(\rm km~s^{-1})$ & $150_{-40}^{+30}$  \\
          & $v_{\rm sys}~(\rm km~s^{-1})$ & $-750_{-80}^{+70}$  \\
          & $A_{\rm N}/A_{\rm N,~solar}$ & $3.7_{-0.8}^{+0.9}$$^*$  \\
{\tt dbb} & $\rm norm~(10^{20}~\rm cm^{2})$ & $8.7_{-0.5}^{+0.5}$  \\
            & $kT~(\rm keV)$ & $10~(f)$  \\
{\tt file} (K96) & $\rm norm~(10^{43}~erg~s^{-1})$ & $1.5_{-0.1}^{+0.1}$  \\
{\tt comt} & $\rm norm~(10^{54}~\rm ph~s^{-1}\, keV^{-1})$ & $1.1_{-4}^{+6}$  \\
            & $kT_{\rm seed}~(\rm keV)$ & $10~(f)$  \\
            & $kT_{\rm plas}~(\rm keV)$ & $0.24_{-0.03}^{+0.04}$  \\
            & $\tau$ & $15_{-2}^{+3}$  \\
{\tt pow} & $\rm norm~(10^{51}~\rm ph~s^{-1}\, keV^{-1})$ & $3.04_{-0.05}^{+0.05}$  \\
          & $\Gamma$ & $1.87_{-0.01}^{+0.01}$$^\dagger$  \\
          & $E_{\rm cut}$ (keV) & $800~(f)$  \\
{\tt gaus} & $\rm norm~(10^{48}~\rm ph~s^{-1})$ & $7.9_{-0.7}^{+0.7}$  \\
          & $E~(\rm keV)$ & $6.43_{-0.02}^{+0.02}$  \\
          & $\sigma~(\rm km~s^{-1})$ & $1300_{-300}^{+300}$  \\
{\tt pexmon} & $\rm norm~(10^{7}~\rm ph~s^{-1}\,cm^{-2}\, keV^{-1})$ & $1.2_{-0.1}^{+0.1}$  \\
          & $\Gamma$ & $1.87_{-0.01}^{+0.01}$$^\dagger$  \\
          & $E_{\rm cut}$ (keV) & $800~(f)$  \\
\hline
\end{tabular}
\end{center}
\hspace{-1.0cm}
\caption{\small{Best-fitting parameters, their values and errors for the model fit to the data from \textit{XMM-Newton} (\textit{OM}, \textit{RGS}, and \textit{pn}), and \textit{NuSTAR}). Parameters labeled with an ``({\it f})'' are held fixed in the fit, while those marked with an ``*'' and ``$\dagger$'' are tied to another parameter (see Sect. \ref{ssec:SED}). Parameters not listed are fixed at their default values. Model components with no free parameters are omitted.}}
\label{tab:sed}
\end{table}
}

{\small
\begin{table}[h]
\hspace{-1.0cm}
\begin{center}
\begin{tabular}{lll}
\hline\hline
{\bf Component} & {\bf Parameter (units)} & {\bf Value} \\
\hline
{\tt comt} & $\rm norm~(10^{55}~\rm ph~s^{-1}\, keV^{-1})$ & $1.3_{-0.2}^{+0.2}$  \\
            & $kT_{\rm seed}~(\rm keV)$ & $10.0~(f)$  \\
            & $kT_{\rm plas}~(\rm keV)$ & $0.240~(f)$  \\
            & $\tau$ & $14.96~(f)$  \\
{\tt pow} & $\rm norm~(10^{51}~\rm ph~s^{-1}\, keV^{-1})$ & $3.14_{-0.08}^{+0.07}$  \\
          & $\Gamma$ & $1.816_{-0.009}^{+0.008}$  \\
{\tt H97*spei} & $ L~(10^{40}~\rm erg~s^{-1})$ & $26_{-2}^{+2}$  \\
            & $r_2~(r_g)$ & $190_{-20}^{+50}$  \\
            & $i~(^\circ)$ & $48.8_{-0.2}^{+0.7}$  \\
            & $q$ & $1.86_{-0.13}^{+0.08}$  \\
{\tt H97*vgau} & $ L~(10^{40}~\rm erg~s^{-1})$ & $8.2_{-0.4}^{+0.4}$  \\
          & $\sigma_v~(\rm km\,s^{-1})$ & $1400_{-100}^{+100}$  \\
{\tt H97*vcom} & $ L~(10^{40}~\rm erg~s^{-1})$ & $0.8_{-0.4}^{+0.5}$   \\
{\tt H97} & $ L~(10^{40}~\rm erg~s^{-1})$ & $1.3_{-0.2}^{+0.3}$  \\

\hline
\end{tabular}
\end{center}
\hspace{-1.0cm}
\caption{\small{Best-fitting parameters, their values and errors for the emission components of the model fit to the \textit{Resolve} time-averaged spectrum. Parameters marked with an {\it ``f''} are held fixed in the fit at the given value, and parameters not listed are fixed at their defaults.}}
\label{tab:emission}
\end{table}
}

\clearpage
\bibliography{n3516abs}{}
\bibliographystyle{aasjournalv7}

\end{document}